\documentclass[conference]{IEEEtran}
\ifCLASSINFOpdf
 
\else

\fi

\usepackage{epsf}
\usepackage{graphics,graphicx,color,multirow,subfigure}
\usepackage{cite}
\usepackage{multirow}
\usepackage{amsmath}
\usepackage[keeplastbox]{flushend}
\usepackage{enumerate}
\usepackage{txfonts}
\usepackage{graphicx}
\usepackage{subfigure}
\usepackage{algorithm}
\usepackage{algorithmic}
\usepackage{tikz}
\usepackage{booktabs}
\usepackage{epsf}
\usepackage{flushend}

\pdfoutput=1
\begin{document}

\title{Temporal Analysis of Transaction Ego Networks with Different Labels on Ethereum}
\author{\IEEEauthorblockN{Baoying Huang\IEEEauthorrefmark{1},
		Jieli Liu\IEEEauthorrefmark{2}, Jiajing Wu\IEEEauthorrefmark{1}\IEEEauthorrefmark{4}, Quanzhong Li\IEEEauthorrefmark{1}, Hao Lin\IEEEauthorrefmark{3}
	}
	\IEEEauthorblockA{\IEEEauthorrefmark{1}School of Computer Science and Engineering, Sun Yat-sen University, Guangzhou 510006, China\\}
	\IEEEauthorblockA{\IEEEauthorrefmark{2}School of Software Engineering, Sun Yat-sen University, Zhuhai 519082, China\\}
	\IEEEauthorblockA{\IEEEauthorrefmark{3}Merchants Union Consumer Finance Company Limited, Shenzhen 518000, China\\}
    \IEEEauthorblockA{\IEEEauthorrefmark{4}Correponding Author: wujiajing@mail.sysu.edu.cn}
	}
	
\maketitle
\begin{abstract}
Due to the widespread use of smart contracts, Ethereum has become the second-largest blockchain platform after Bitcoin. Many different types of Ethereum accounts (ICO, Mining, Gambling, etc.) also have quite active trading activities on Ethereum. Studying the transaction records of these specific Ethereum accounts is very important for understanding their particular transaction characteristics, and further labeling the pseudonymous accounts. However, traditional methods are generally based on static and global transaction networks to conduct research, ignoring useful information about dynamic changes. Our work chooses six kinds of important account labels, and builds ego networks for each kind of Ethereum account. We focus on the interaction between the target node and neighbor nodes with temporal analysis. Experiments show that there is a significant difference between various types of accounts in terms of several network features, helping us better understand their transaction patterns. To the best of our knowledge, this is the first work to analyze the dynamic characteristics of Ethereum labeled accounts from the perspective of transaction ego networks.
\end{abstract}

\IEEEpeerreviewmaketitle

\section{INTRODUCTION}

%区块链 ->以太坊 -> 以太坊交易网络和现有的研究,介绍研究motivation介绍我们的研究问题和contribution

%区块链
After the development of more than a decade, blockchain technology has become a buzzword and brought world-shaking innovations to human life. Generally speaking, blockchains are append-only distributed ledgers originated from Bitcoin~\cite{nakamoto2008bitcoin}. Transactions on blockchain platforms are publicly verifiable based on open data architecture. Ethereum~\cite{wood2014ethereum} is the largest public blockchain enabling Turing-complete smart contracts, which are computer programs that can automatically be executed in the Ethereum Virtual Machine (EVM). Since the outbreak of COVID-19, investors have found that Ethereum is suitable as a short-term safe-haven~\cite{mnif2020cryptocurrency}.
%Ethereum~\cite{wood2014ethereum} is the largest public blockchain enabling %Turing-complete smart contracts, which are computer programs that can %automatically be executed in the Ethereum Virtual Machine (EVM) according to %the pre-defined program logic. Especially since the outbreak of COVID-19, %investors have found that Ethereum is suitable as a short-term %safe-haven~\cite{mnif2020cryptocurrency}. 
%
%With a combination of peer-to-peer transmission, cryptography, and consensus protocol, blockchains provide a distributed environment for cryptocurrency transactions without any third party.
%The introduction of smart contracts paves a new way for the application of blockchain technology, and a wide variety of decentralized applications (DApps) are constantly emerging.
% 以太坊

%As a result, the market price and trading volume of Ethereum continued to increase.
Therefore, increasing research efforts have been devoted to the analysis of the publicly accessible Ethereum transaction records. A considerable part of existing work modeled the transaction data into networks, which are a general language in describing interacting systems\cite{victor2019measuring}\cite{mixingdetectionjieli}~\cite{YuxuanSMC}. In the Ethereum transaction network, an Ethereum account is regarded as a node, and the edges between two nodes present the interactions between the two accounts with attributes such as transaction value, timestamp and so on.
% In Ethereum, the transfer of value and information between accounts can be straightforwardly represented as the interactions in a network system. 

% related work 和 motivation
Recent research~\cite{lin2020modeling}~\cite{WU2021103139}~\cite{WhoAreThePhisher} have analyzed the entire Ethereum transaction network from a static perspective for understanding the whole Ethereum ecosystem. They have built several static networks using transaction data, and analyzed network properties such as density and clustering coefficients, comparing them with social networks and the Web. However, the daily new transaction volume of Ethereum is quite large, and static analysis leads to the loss of insights from a dynamic evolution perspective. Consequently, several recent studies~\cite{geng2021vw}~\cite{LinDanTCSS}~\cite{zhao2021temporal} provided insights into the temporal and evolution properties of the entire Ethereum transaction network. But there is few studies investigating the microscopic structure of a kind of nodes in the Ethereum transaction network, which can give explanations for the different behaviors of accounts. Analyzing the Ethereum transaction network from a perspective of various types of nodes can provide us with a new understanding of their microscopic network structure, which also helps to label the pseudonymous accounts in Ethereum. 
% Especially, different kinds of labeled accounts play various roles in the Ethereum transaction network.
% 本文的工作
In this paper, we are interested in studying the temporal changes of multiple transaction characteristics based on the ego network of each labeled account. The ego network consists of the central node ``EGO'' and the nodes directly connected to EGO (these are called ``ALTER''), as well as the connections between the EGO and ALTER nodes, ALTER and ALTER nodes~\cite{odella2006using}. Ego network is widely used in social network analysis and helps to find many patterns~\cite{leskovec2008microscopic}~\cite{arnaboldi2012analysis}. We mainly select six typical Ethereum account labels, including four normal labels (ICO, Mining, Gambling, Exchange) and two illegal labels (Ponzi and Phish), which are collected from Etherscan's label word cloud\footnote{https://cn.etherscan.com/labelcloud}. A normal account has no illegal activities such as deceiving users. While an illegal account is the opposite. Then, we collect the required transaction records from a blockchain data platform and build multiple ego networks. After analyzing how the transaction features change over time for each type of node, we discover the different features between various types of accounts and better understand their roles on Ethereum.

% 章节的安排
The remainder of this paper is structured as follows. Section~\ref{sec:dataset} introduces the process of collecting transaction data and building the ego network of various accounts. The analysis of network properties will be given in Section~\ref{sec:statistics}. Finally, we will give a brief conclusion in Section~\ref{sec:conclusion}.

% 数据收集和处理部分

\section{DATASET AND EXPERIMENT SETUP\label{sec:dataset}}
%收集的数据类型、来源、大小，网络的构建和实验配置
In this part, we firstly introduce the source and processing of data, and then explain the definition of ego network constructed by using transaction data.

% 数据的收集
\subsection{Data Collection}
Etherscan is one of the most widely used Ethereum browsers, recording lots of account labels that are reliably provided by multiple users. A lot of previous research work on Ethereum also used the labeled accounts on Etherscan~\cite{yuan2020detecting}~\cite{chen2020phishing}. We choose six types of the most common and important accounts on Ethereum as the targets. Especially, according to the crime report about cryptocurrency~\cite{chainalysis}, ponzi schemes and phish scams have defrauded millions of dollars. Now we will introduce the definition of these accounts in detail.
\begin{itemize}
\item[$\bullet$] $\textbf{ICO}$: The accounts used to receive crowdfunding for the initial offering of the tokens
\item[$\bullet$] $\textbf{Mining}$: The accounts of miners who participate in block mining
\item[$\bullet$] $\textbf{Gambling}$: The accounts of the gambling platforms
\item[$\bullet$] $\textbf{Exchange}$: The accounts of exchanges that allow users to sell or buy cryptocurrencies
\item[$\bullet$] $\textbf{Ponzi}$: The accounts related to the manipulator of ponzi schemes
\item[$\bullet$] $\textbf{Phish}$: The accounts related to the manipulator of phish scams
\end{itemize}

In addition, we collect all target labeled accounts as well as their complete $Ether$\footnote{$Ether$ is the official cryptocurrency of Ethereum.} transaction records. In order to construct the ego network, we also collect the transactions between their first-order neighbors. These transaction records are extracted from the latest Ethereum dataset on XBlock\footnote{http://xblock.pro/xblock-eth.html}, which is a blockchain data platform. Therefore, a total of 2584 accounts and nearly 38 million transaction records are collected.

% 网络构建
\subsection{Construction of Ego Networks}

%Through the above operations on the Ethereum data, we obtain a transaction network. 
Let $ G =(V, E) $ represents the ego network of an Ethereum account, where $V$ represents the set of nodes, and $E$ is the set of edges. $V$ contains the central node ``EGO'' and its neighbor nodes ``ALTER''. The transactions between EGO and ALTER nodes, ALTER and ALTER form the edge set $E$. Particularly, an edge can be represented as $e= <u, v, w, t>$, indicating that at time $t$, account $u$ transfers Ether of amount $w$ to account $v$. Consequently, the ego network in our work is a weighted directed network.

\begin{figure}[htbp]
    \centering
    \includegraphics[width=7cm]{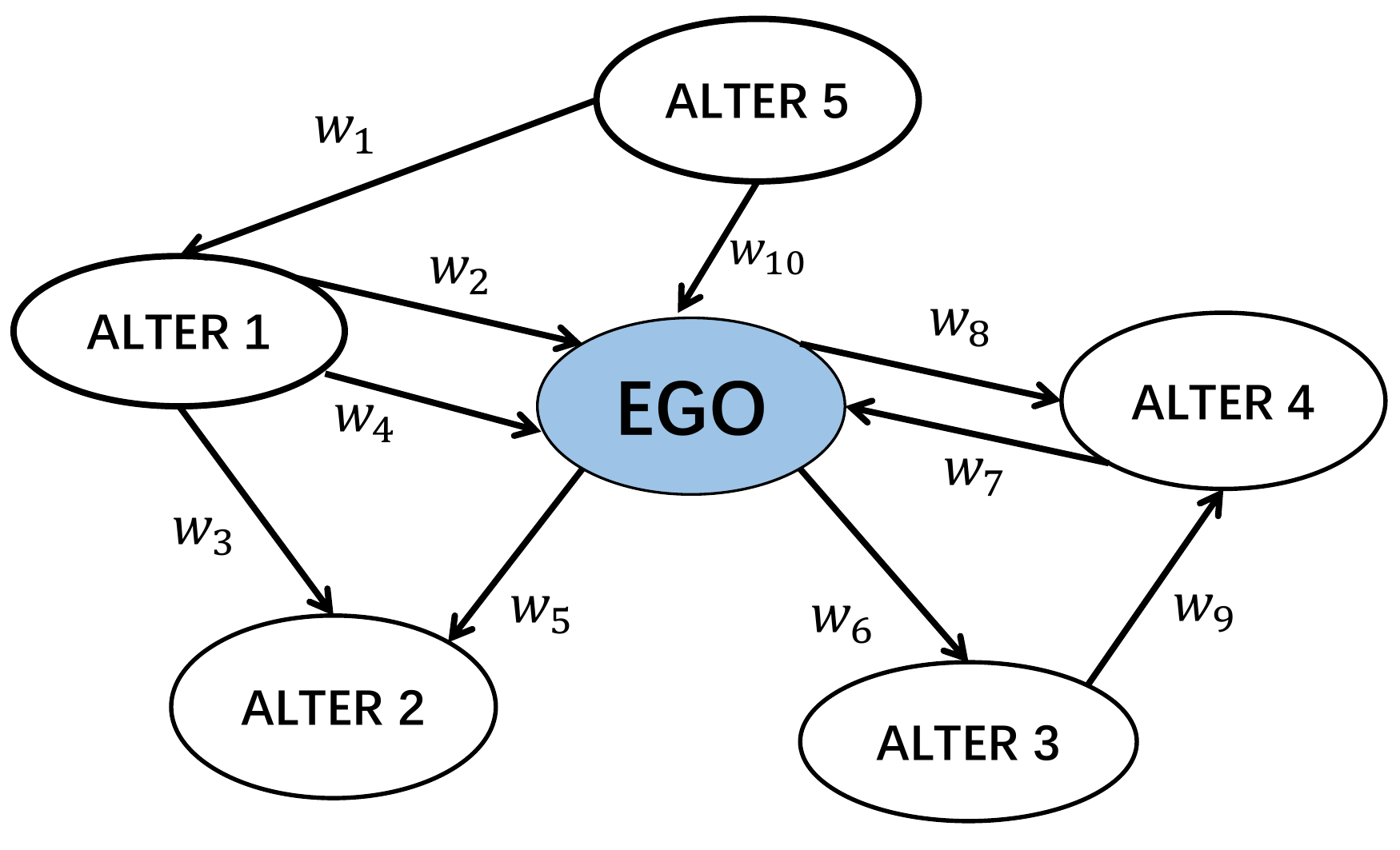}
    \caption{An example Of ego network.}
    \label{fig:ego_network}
\end{figure}

Fig.~\ref{fig:ego_network} shows an example of an ego network. Moreover, we construct an ego network for each labeled account, classify the ego networks of the same label into one category, and take the average of various statistics as the final feature value of the label accounts. The details of each feature will be shown in the next section.

% 插入ego network图片

\section{THE ANALYSIS OF NETWORK PROPERTIES\label{sec:statistics}}
% 介绍交易特征
First, we focus on the local clustering coefficient of each labeled account to determine whether the nodes in the network are closely connected. Then we analyze several transaction characteristics of labeled accounts. Finally, we obtain the unique transaction features of each labeled account, which also explain the role of the corresponding label in the Ethereum ecosystem.
% 子网的局部聚类系数
\subsection{Local Clustering Coefficient}
The local clustering coefficient is a measure of the degree to which the neighbors of the central node in the graph tend to cluster together. Let $G_i = (V, E)$ represents the ego network whose ego is node $i$. Here the edge of node $u$ pointing to node $v$ is written as $e_{uv}$. Then $N_i$ represents the neighbor nodes of node $i$ and $ k_i = |N_i| $. Hence the local clustering coefficient of node $i$ is:
\begin{equation}\label{localcoeff}
\tau_i = \frac{|\{e_{uv}:u, v\in N_i,e_{uv} \in E\}|}{k_i(k_i - 1)}.
\end{equation}

Then we calculate the average local clustering coefficient of each type of ego network, and the results are shown in Table I. According to the table, the probability of transactions occurring between the neighbors of ICO accounts is about 18$\%$, followed by Mining accounts and Exchange accounts. It tells us there is a higher probability that investors of ICO accounts will have transactions with each other. The coefficient of Gambling accounts is the smallest, which means the users participating in the gambling are generally not connected. Besides, the local clustering coefficient of Phish accounts is larger than that of Ponzi accounts, indicating that the neighbors of Phish accounts are also more likely to be connected. We speculate that there are fixed fraud partners in phishing activities.

\begin{table}[ht]
\centering  
\caption{The local clustering coefficient of labeled egos}
\label{table1}  %表格名字，用于正文中引用表格
\setlength{\tabcolsep}{2mm}
\begin{tabular}{|c|c|c|c|c|c|c|}
\hline
Label & ICO & Mining & Gambling & Exchange & Ponzi & Phish \\ 
\hline
$\tau$ & 0.1812  & 0.1354 & 0.0242 & 0.1324 & 0.0379 & 0.1226 \\
\hline
\end{tabular}
\end{table}

% 节点的生命周期
\subsection{Ego's Life Cycle and Time Window}
For an Ethereum account $u$ with $M$ transactions, we define its life cycle as $Life_u = t_M - t_1$, where $t_M$ represents the time when the last transaction occurred and $t_1$ represents the time when the first transaction occurred. We count the life cycles of each kind of account and record its median and maximum values in Table II. It can be concluded that the life cycle of illegal accounts is much shorter than that of normal accounts, and their life cycle is less than 20 days. This is because the criminals usually transfer the illegal gains out of the accounts as soon as possible and no longer use the original accounts to enhance anonymity.
% Because the criminals transfer the illegal gains as soon as possible before the fraud is detected, and they usually abandon the accounts and no longer use it. 
% However, the normal account will be used all the time, so its life is longer.
%3. Gambling account 0x777f415324d56e1d54fa832902d8797db7a4c57c, longest life cycle is the account of the Ethernet fang 1 xbet platform (an online gambling company), are now trading, has repeatedly suspected of illegal scams.
%生命周期的对比还没写
\begin{table}[t]
\centering  
\caption{The life cycle of labeled ego (days)}
\label{table2}  %表格名字，用于正文中引用表格
\setlength{\tabcolsep}{1.4mm}
\begin{tabular}{|c|c|c|c|c|c|c|}
\hline
Label & ICO & Mining & Gambling & Exchange & Ponzi & Phish \\ 
\hline
Median & 419.92  & 646.24 & 292.59 & 603.06 & 19.99 & 15.76 \\
\hline
Mean & 1700.18  & 2150.09 & 1527.35 & 2132.30 & 1642.22 & 1529.09 \\
\hline
\end{tabular}
\end{table}

%exchange和phish的生命周期分布对比

%\begin{figure}[htbp]
%    \centering
%    \includegraphics[width=8.5cm]{iscas_yq/figures/phish-exchange.pdf}
%    \caption{The distribution of two types of Ego's life}
%   \label{fig:cmp_life}
%\end{figure}

Afterwards, in order to observe the ego network of each type of label from a dynamic perspective, we refer to the previous work~\cite{zhao2021temporal} and consider two types of time windows, the sliding window and the incremental window. Firstly, we take $1/5$ of the life cycle of each type of labeled account as a fixed time span $T$, which is also the initial length of the two time windows. The sliding distance of the sliding window is $T$ each time, while the size of the incremental window increases by $T$. Next, we use different time windows to extract transaction details of the egos, and obtain different characteristics in five phases.

% 交易次数
\subsection{The number of transactions between EGOs and ALTERs}
\begin{figure}[t]
    \includegraphics[width=8.5cm]{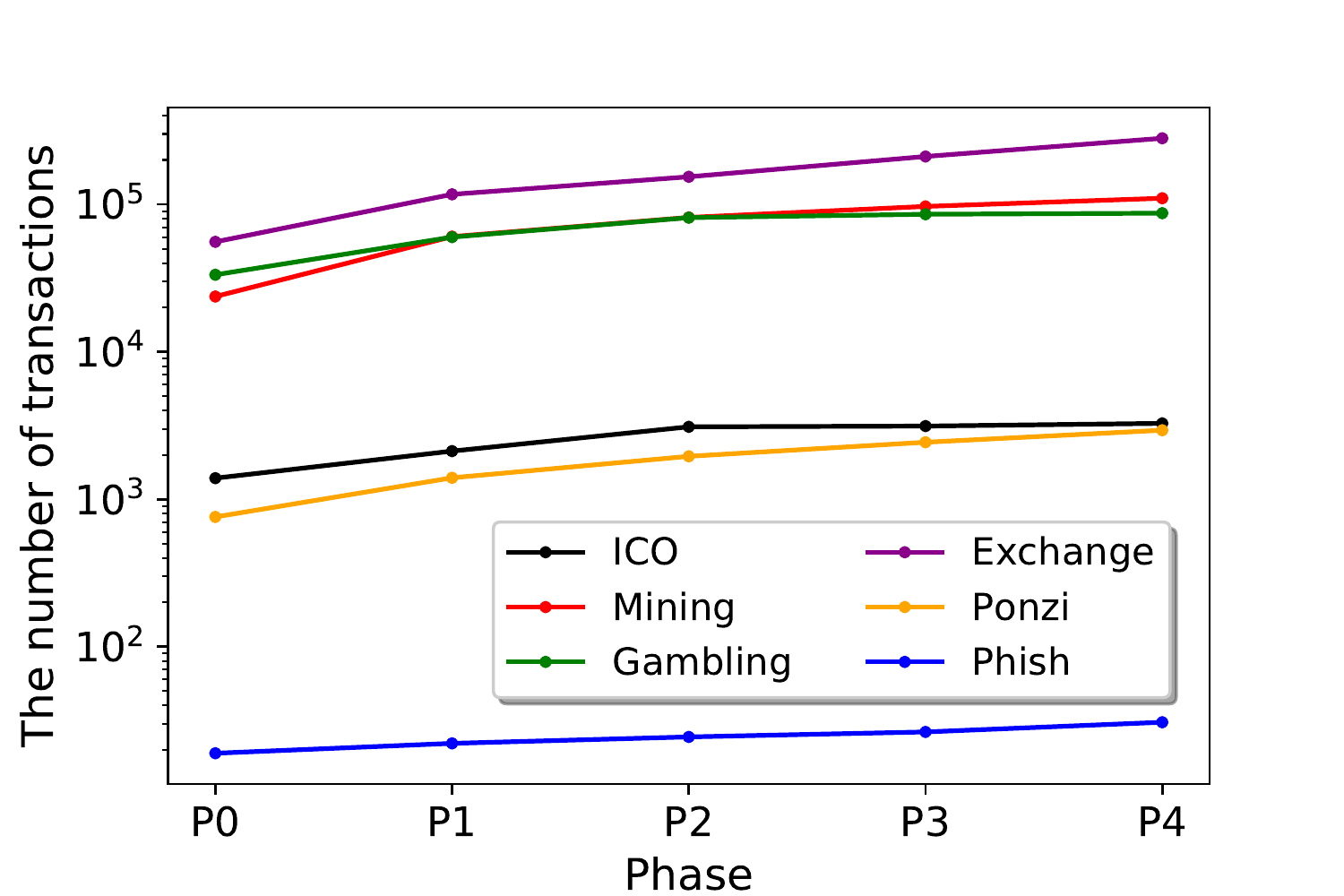}
    \caption{Changes of the number of transactions at each phase}
    \label{fig:alltx}
\end{figure}

In order to study the transaction activities of different types of accounts, we compare the evolution in the number of transactions between EGO and ALTER nodes during their life cycles in the incremental window mode. 
It can be seen from Fig.~\ref{fig:alltx} that the number of transactions of illegal accounts is less than that of the normal accounts. In addition, the number of transactions on Exchange, Mining and Gambling accounts is relatively large. Because Exchange accounts are an intermediary on Ethereum, they generate a lot of transactions with users to buy and sell tokens. According to the description on Etherscan, we can know that Mining accounts are mining pools composed of many miners. So they usually return bonuses to miners multiple times. Additionally, Gambling accounts representing a kind of entertainment activity, also attract lots of users. Significantly, the curves of ICO accounts and Ponzi accounts are relatively similar, which shows that they have certain commonalities in the trading pattern. We presume that it is one of the reasons why it's difficult for users to distinguish whether the investment project is a scam or not.
%Among them, the transaction number of Exchange accounts and Mining accounts grows rapidly, indicating that they both have high reliability and quite frequent transactions.
% 插入all tx图片

For the purpose of further exploring the roles of the labeled accounts on Ethereum, the changes in the ratio of two kinds of transactions (in/out) in the incremental window mode are also recorded. An in-transaction is an interaction in which the EGO receives Ether, and an out-transaction is an interaction in which the EGO transfers Ether to another account.
\begin{table}[t]
\centering  
\caption{Changes in the proportion of the number of in/out transactions}
\label{table3}  %表格名字，用于正文中引用表格
\setlength{\tabcolsep}{1.8mm}
\begin{tabular}{|c|c|c|c|c|c|c|}
\hline
   & Phase & P0 & P1 & P2 & P3 & P4 \\ 
\hline
\multirow{2}*{ICO} & in & 92.36$\%$ & 90.83$\%$ & 62.84$\%$ & 62.92$\%$ & $\textbf{61.24$\%$}$ \\
\cline{2-7}
~  & out & 7.64$\%$ & 9.17$\%$ & 37.16$\%$ & 37.08$\%$ & 38.76$\%$ \\
\hline
\multirow{2}*{Mining} & in & 0.19$\%$ & 0.14$\%$ & 0.17$\%$ & 0.17$\%$ & 0.16$\%$ \\
\cline{2-7}
~  & out & 99.81$\%$ & 99.86$\%$ & 99.83$\%$ & 99.83$\%$ &  $\textbf{99.84$\%$}$ \\
\hline
\multirow{2}*{Gambling} & in & 55.73$\%$ & 60.13$\%$ & 61.45$\%$ & 61.31$\%$ & $\textbf{61.27$\%$}$ \\
\cline{2-7}
~  & out & 44.27$\%$ & 39.87$\%$ & 38.55$\%$ & 38.69$\%$ & 38.73$\%$ \\
\hline
\multirow{2}*{Exchange} & in & 43.83$\%$  & 44.18$\%$ & 41.34$\%$ & 33.25$\%$ & 26.67$\%$ \\
\cline{2-7}
~  & out & 56.17$\%$ & 55.82$\%$ & 58.66$\%$ & 66.75$\%$ & $\textbf{73.33$\%$}$ \\
\hline
\multirow{2}*{Ponzi} & in & 29.04$\%$ & 28.28$\%$ & 28.38$\%$ & 28.98$\%$ & 30.58$\%$ \\
\cline{2-7}
~  & out & 70.96$\%$ & 71.72$\%$ & 71.62$\%$ & 71.02$\%$ & $\textbf{69.42$\%$}$ \\
\hline
\multirow{2}*{Phish} & in & 61.15$\%$ & 61.82$\%$ & 61.46$\%$ & 61$\%$ & $\textbf{57.78$\%$}$ \\
\cline{2-7}
~  & out & 38.85$\%$ & 38.18$\%$ & 38.54$\%$ & 39$\%$ & 42.22$\%$ \\
\hline
\end{tabular}
\end{table}

%转入转出金额对比
\begin{figure*}[t]
\centering
\subfigure[ICO]{
\includegraphics[width=6.3cm]{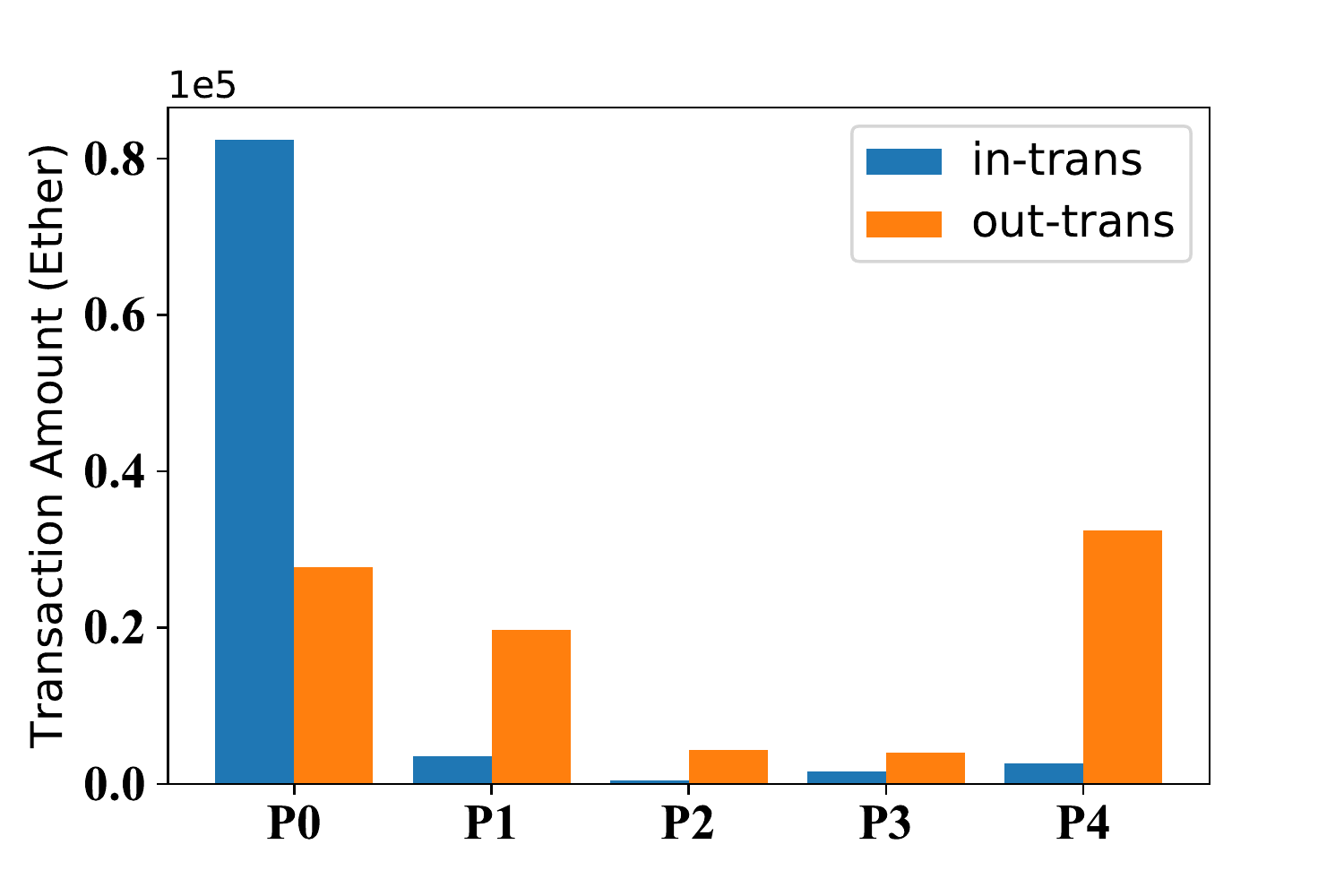}
\hspace{-14.5mm}
%\caption{fig1}
}
\hfill
\subfigure[Mining]{
\centering
\includegraphics[width=6.3cm]{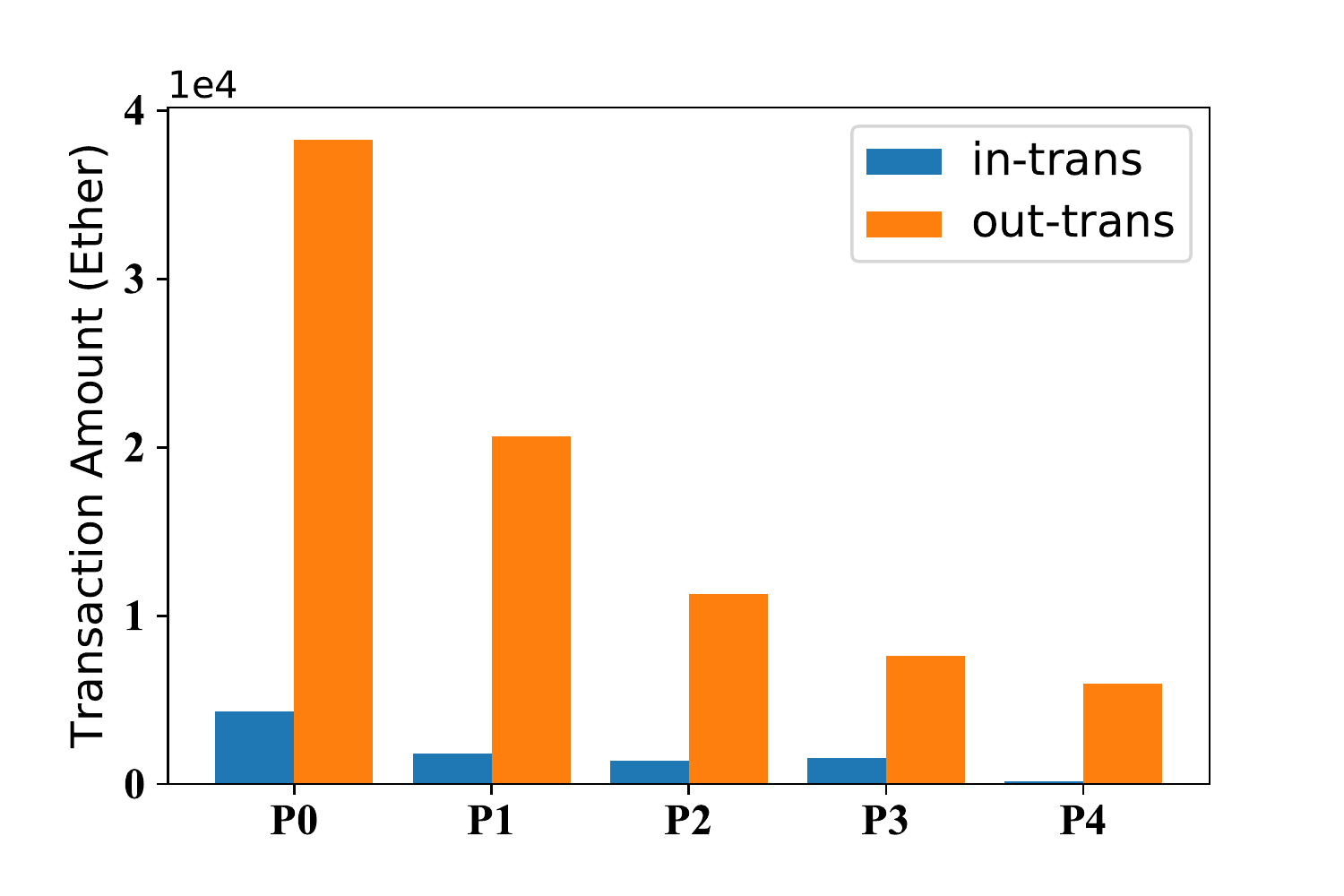}
\hspace{-14.5mm}
}
\hfill
\centering
\subfigure[Gambling]{
\includegraphics[width=6.3cm]{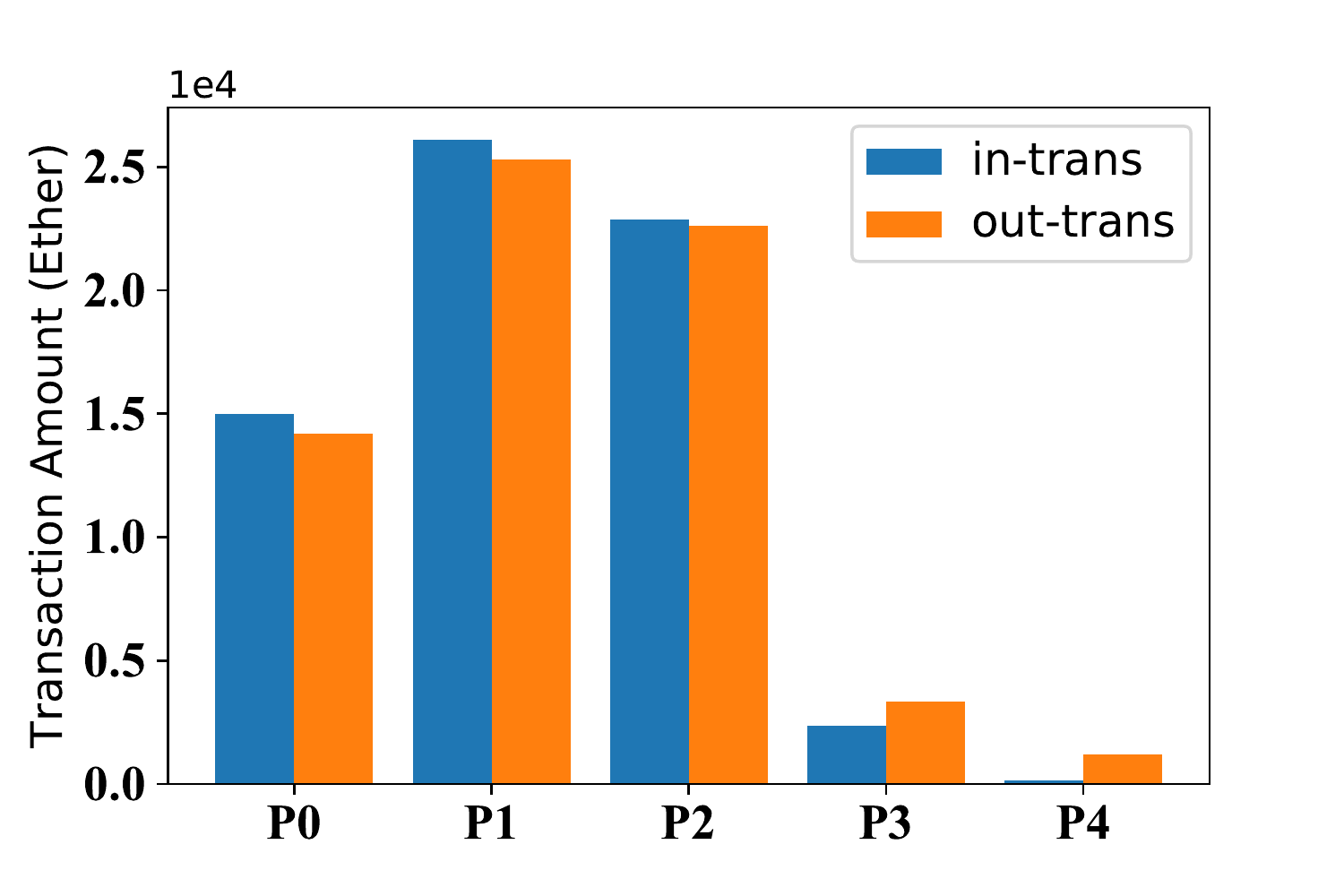}
}
\hfill
\centering
\subfigure[Exchange]{
\includegraphics[width=6.3cm]{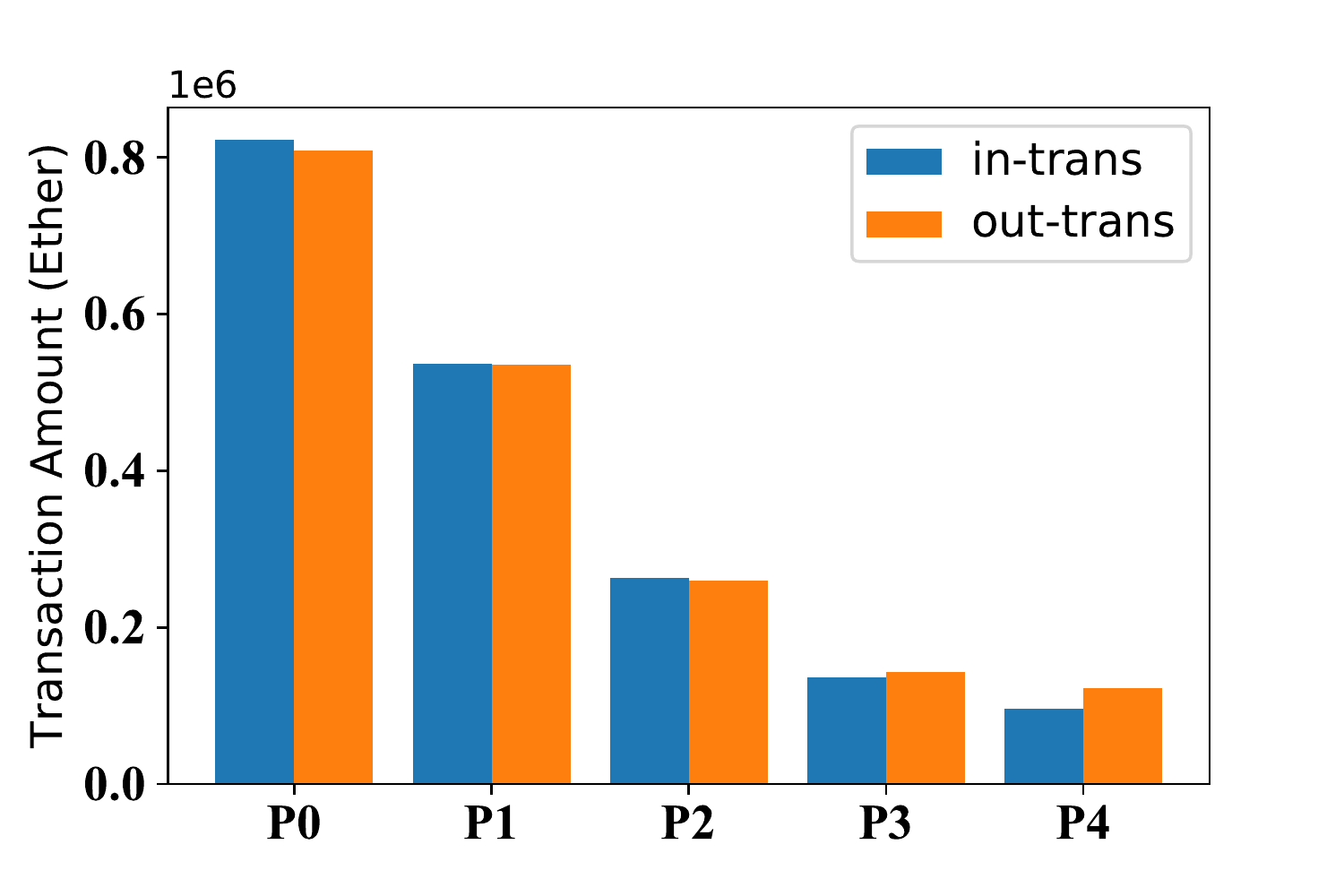}
\hspace{-14.5mm}
}
\hfill
\centering
\subfigure[Ponzi]{
\includegraphics[width=6.3cm]{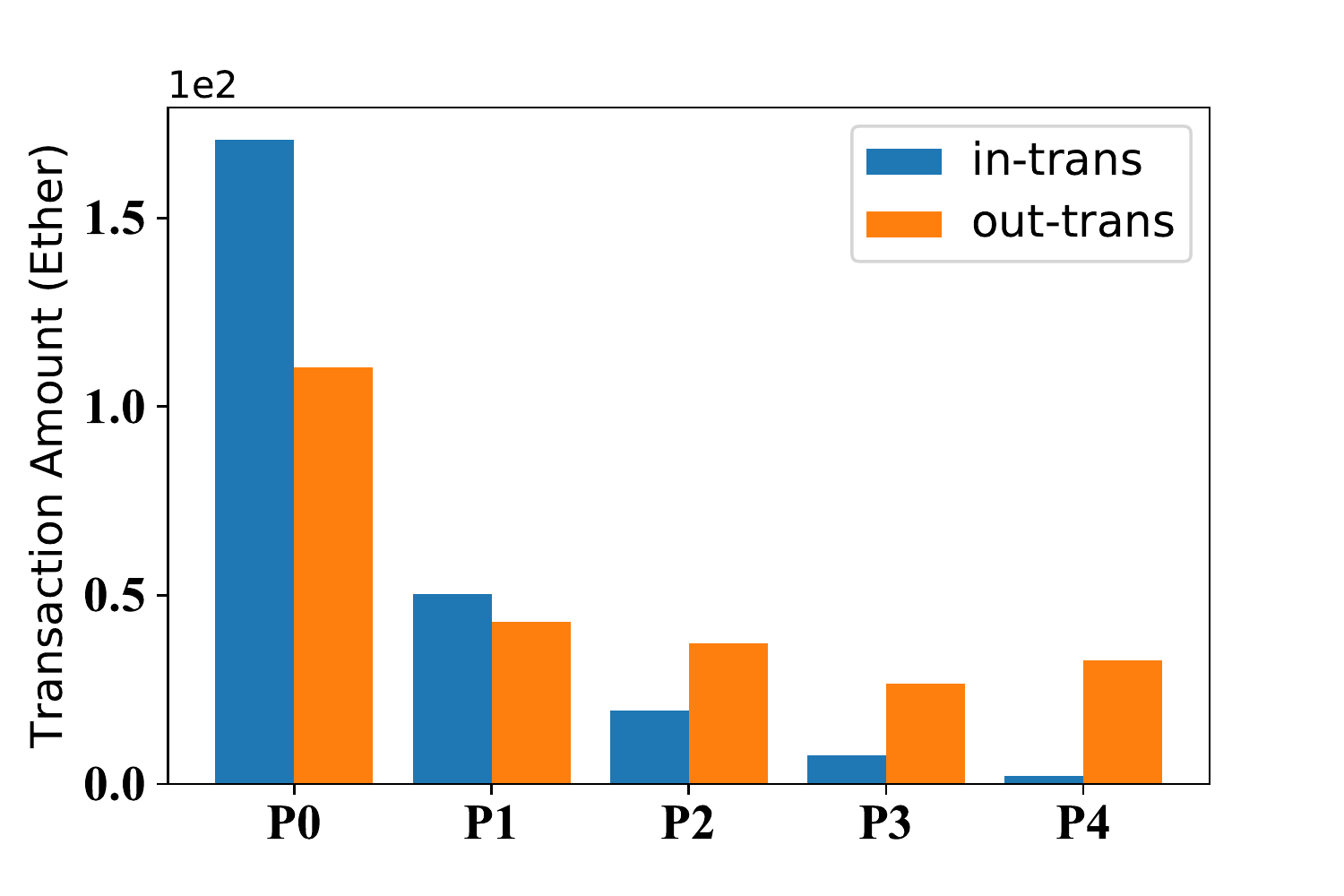}
\hspace{-14.5mm}
}
\hfill
\centering
\subfigure[Phish]{
\includegraphics[width=6.3cm]{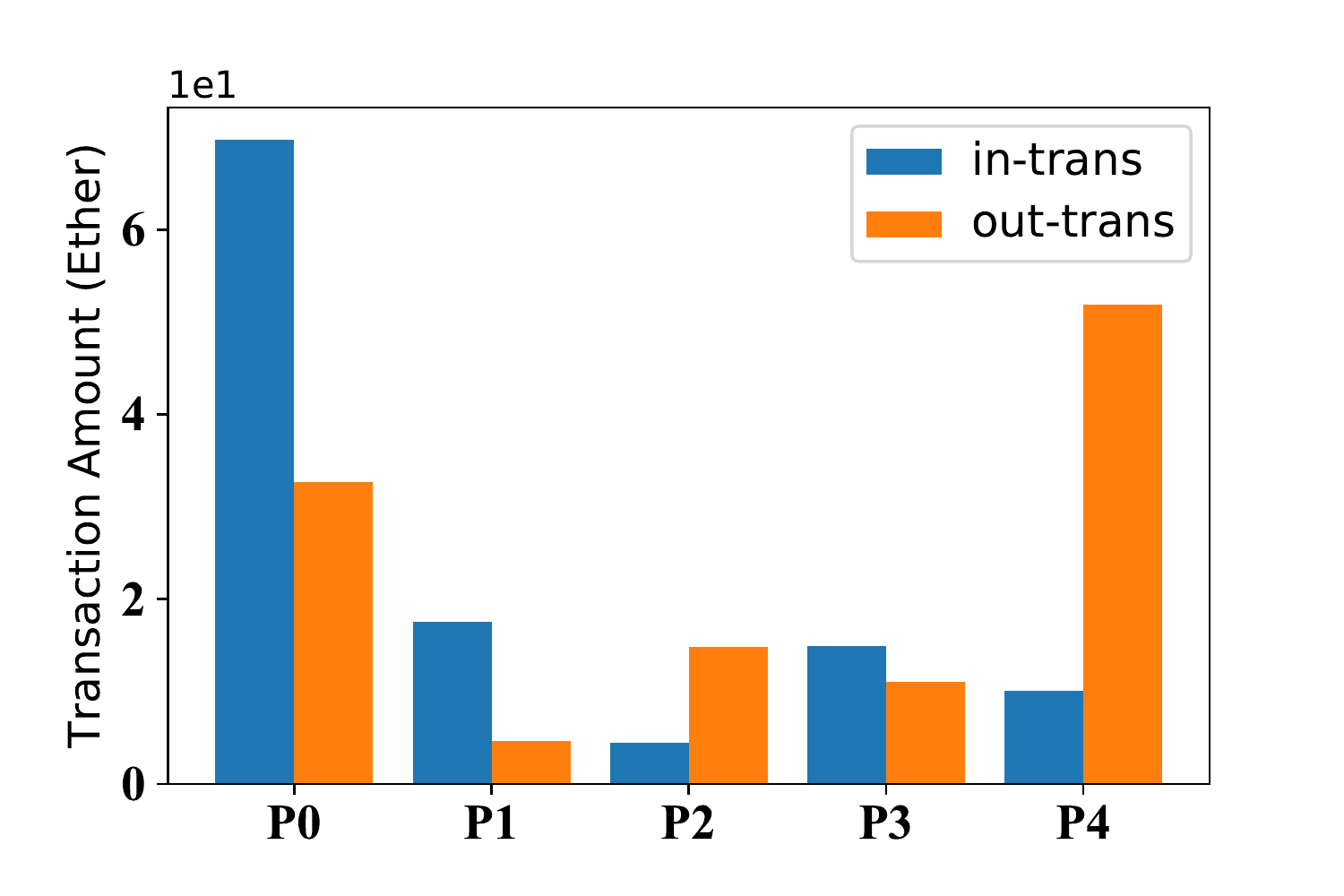}
}
\caption{Changes of the transaction amount at each phase}
\label{fig:txamount}
\end{figure*}

The results in Table III show that only Mining, Exchange, and Ponzi accounts have a higher proportion of out-transactions than in-transactions. We suppose that the reasons for this phenomenon are as follows. Firstly, the rewards obtained from mining blocks will be distributed to users who invest in miners, which is a long-term stable mechanism. Secondly, when an Exchange account is used as the initiator of transactions, it means that users are buying Ether from this platform. Generally speaking, investors will continue to purchase Ether multiple times and then sell them all at the highest price point. And thus  Exchange accounts have more in-transactions than out-transactions. Thirdly, a Ponzi account intends to act like a normal project at first to attract more people. Then, a small part of the investment of latecomers will also be used as bonuses to return to the initial users several times. 

Moreover, the in-transaction ratio of ICO accounts is relatively large at the initial phase of token issuance, and the later transaction volume gradually decreases. The other two kinds of accounts with a higher proportion of in-transactions are Gambling and Phish. A gambling activity is a game where the dealer makes a profit, so the number of users who win money is less than the total number of participants. Phish accounts are fraudulent, so they deceive multiple users to obtain Ether, and then transfer out all the illegal gains at the final phase.

% 交易金额
\subsection{The amount of transactions between EGO and ALTER}
Besides the number of transactions, the transaction amount is also an extremely important feature. Therefore, we observe the transfer amount of Ethereum accounts at each phase in the sliding window mode, and analyze the difference and essence.
It can be seen from Fig.~\ref{fig:txamount} that ICO accounts initially receive a large amount of investment and then gradually transfer it to the token-related accounts. Interestingly, the total amount transferred out from Mining accounts is always greater than the total amount transferred in, because most of the balance in the accounts comes from mining rewards. These rewards are not be recorded in the transactions, but they can be queried through the Ethereum browser. In addition, when no player participates in gambling games, Gambling accounts will transfer the balance out, indicating that Gambling accounts are essentially a profitable entity. 
In general, only the transfer-in and transfer-out amount at all phases of Exchange accounts are basically the same, which also reflects the role of Exchange accounts as an intermediary for the sale of tokens. Finally, we find that both types of illegal accounts have reaped huge profits, which means it's urgent to identify and combat fraudulent accounts on Ethereum. It is worth noting that Phish accounts usually transfer out a large amount of Ether in the final phase. While Ponzi accounts transfer Ether out in batches, and some Ether will be returned to investors.

\section{CONCLUSIONS AND FUTURE WORK\label{sec:conclusion}}

With the rapid development of Ethereum, a large number of accounts on Ethereum have involved in various business roles. For exploring the transaction behavior of these accounts and obtaining a more comprehensive description of their roles in the Ethereum ecosystem, we used transaction records to construct multiple ego networks with transaction amount and time information for each type of account. By analyzing the dynamic changes of the interaction characteristics of EGO (labeled accounts) and ALTER (first-order neighbor accounts), we found that the trading of Mining accounts and Exchange accounts is more stable. In addition, the two illegal accounts, Ponzi and Phish, also have obvious abnormalities in the ratio of the amount of in/out transactions. The dynamic changes of transaction characteristics also enabled us to label unknown accounts. Most of all, we understood the evolutionary phases of scams, which can provide some help in identifying fraudulent activities on Ethereum.

In the future, we will continue to expand our research by investigating more network properties in the dynamic transaction network, such as transaction objects and network motifs. We expect the corresponding results to tell us  whether there is a specific transaction rule and help us discover hidden communities in the network. Furthermore, we will consider more account labels, and more dynamic changes in complex network characteristics.

\section*{Acknowledgment}
The work described in this paper is supported by the National Key R\&D Program of China (2020YFB1006005), the National Natural Science Foundation of China (61973325) and the Natural Science Foundations of Guangdong Province (2021A1515011661)
%\printbibliography

\bibliographystyle{IEEEtran}
\bibliography{iscas_yq.bib}

\end{document}